\documentclass{INTERSPEECH2023}
\usepackage{tabularx,lipsum,xcolor}
\usepackage{amsmath}
\usepackage{multicol, multirow, booktabs, makecell}
\usepackage{cite}
\newcolumntype{Y}{>{\centering\arraybackslash}X}
\newcommand{\newpara}[1]{\vspace{1pt}\noindent\textbf{#1}}

\interspeechcameraready 

\title{Towards single integrated spoofing-aware speaker verification embeddings}
\name{Sung Hwan Mun$^{1,*}$\thanks{$^*$Equal contribution.}, Hye-jin Shim$^{2,*}$, Hemlata Tak$^{3,*}$, Xin Wang$^4$, Xuechen Liu$^{2,5}$, Md Sahidullah$^{6}$, Myeonghun Jeong$^1$, Min Hyun Han$^1$, Massimiliano Todisco$^3$, Kong Aik Lee$^7$, Junichi Yamagishi$^4$, Nicholas Evans$^3$, Tomi Kinnunen$^2$, Nam Soo Kim$^1$, and Jee-weon Jung$^{8,\dag}$\thanks{$^\dag$ Work done while author was in Naver Corporation, South Korea.}}
\address{ 
  $^1$Seoul National University, South Korea,
  $^2$University of Eastern Finland, Finland,\\
  $^3$EURECOM, France,
  $^4$National Institute of Informatics, Japan,
  $^5$Inria, France,
  $^6$TCG CREST, India, \\
  $^7$Institute for Infocomm Research, A*STAR, Singapore,
  $^8$Carnegie Mellon University, USA}
  
\email{shmun@hi.snu.ac.kr, hyejin.shim@uef.fi, tak@eurecom.fr, jeeweonj@andrew.cmu.edu}

\begin{document}
\maketitle
\begin{abstract}
This study aims to develop a single integrated spoofing-aware speaker verification (SASV) embeddings that satisfy two aspects.
First, rejecting non-target speakers' input as well as target speakers' {\em spoofed} inputs should be addressed. 
Second, competitive performance should be demonstrated compared to the fusion of automatic speaker verification (ASV) and countermeasure (CM) embeddings, which outperformed single embedding solutions by a large margin in the SASV2022 challenge.
We analyze that the inferior performance of single SASV embeddings comes from {\em insufficient amount of training data} and {\em distinct nature of ASV and CM tasks}.
To this end, we propose a novel framework that includes multi-stage training and a combination of loss functions. 
Copy synthesis, combined with several vocoders, is also exploited to address the lack of spoofed data.
Experimental results show dramatic improvements, achieving an SASV-EER of 1.06\% on the evaluation protocol of the SASV2022 challenge.
\end{abstract}

\section{Introduction}
While today's state-of-the-art automatic speaker verification (ASV) systems~\cite{kinnunen2010overview,bai2021speaker} offer exceptional reliability across diverse scenarios and domains~\cite{zhang2022mfa,chen2022large}, their vulnerabilities to spoofing and deepfake attacks are well acknowledged.
Such malicious attacks can be generated using text-to-speech (TTS) and voice conversion (VC) algorithms, for instance, in order to deceive an ASV system and provoke an increased rate of false positives. As a result, the development of spoofing and deepfake detection systems, known as countermeasures (CMs), has attracted considerable research interest in recent years~\cite{kinnunen2017asvspoof,wang2020asvspoof,yamagishi21_asvspoof}.

The majority of CM solutions take the form of binary classifiers.
Their role is to protect the ASV system by determining whether a given utterance is bona fide (i.e., genuine) or spoofed.
While the development of such separate CM and ASV subsystems is common to the majority of related research, the recent Spoofing-Aware Speaker Verification (SASV) initiative~\cite{jung22c_INTERSPEECH} was formed in 2022 to promote the exploration of jointly optimized CM and ASV solutions as well as integrated approaches whereby the same tasks are performed by a single classifier.  

The initiative was well supported and attracted the registration of 53 participants. 
While a majority of participants utilized fusion-based approaches, encouragingly, some participants explored end-to-end (E2E) approaches and the training of single, integrated systems~\cite{shim2020integrated,li2020joint,todisco2018integrated}.
E2E approaches involve the training of an SASV model to extract \emph{SASV embeddings} directly (similar to speaker embeddings for an ASV task) without optimization of separate ASV and CM models.
These, however, significantly under-perform solutions based upon the fusion of independent, pre-trained CM and ASV sub-systems at embedding, score, or decision levels~\cite{zhang2022probabilistic, lin2022clips, dku2022sasv, hyu_sasv_2022,kang2022end, teng2022sa}.

Even if their performance is currently behind that of their more traditional counterparts, single, integrated systems are not without appeal. 
A single classifier might be more efficient than two, depending on their relative complexity.
Both CMs and ASV subsystems have potential to deflect spoofing attacks~\cite{ge2022joint}, and jointly-optimized solutions are gaining ground on separate, independently optimized alternatives~\cite{Alenin2022A,dku2022sasv,Zhang2022Norm,zhang2022flyspeech}. Single classifier solutions might also help to avoid the more costly joint optimization or adaptation of dual classifier solutions which might be applied periodically to protect reliability in the face of newly emerging threats. 
Finally, given that they have received far less research attention than their dual classifier counterparts, it is sensible to assume that there might be as-yet untapped potential to improve the competitiveness of single classifier solutions with further research investment.  
We believe that single, integrated solutions still merit attention.

To this end, we propose a novel multi-stage framework for training single SASV embedding extractors.
The framework employs different stages to leverage vast amounts of data, including VoxCeleb2~\cite{chung2018voxceleb2} as well as copy synthesis (CS) data~\cite{Wang2023a}.
Furthermore, considering the characteristics of SASV task, involving two different tasks in nature, delicate optimization using diverse combinations of loss functions is explored to go beyond the simple binary classification. 
We conducted extensive experiments using two recent speaker embedding extractors, namely a selective kernel attention-based time delay neural network (SKA-TDNN)~\cite{mun2023frequency} and a multi-scale feature aggregation conformer (MFA-Conformer)~\cite{zhang2022mfa}.
Together, these contributions narrow significantly the performance gap between embedding-level, fusion-based and single, integrated SASV solutions. \footnote{Code and models are available in~\url{https://github.com/sasv-challenge/ASVSpoof5-SASVBaseline}.}.


\vspace{-0.3cm}
\section{Proposed SASV framework}
Previous studies in SASV have shown the improvements to performance that can be found from using large-scale datasets~\cite{jung22c_INTERSPEECH, zhang2022probabilistic, Zhang2022Norm, kang2022end, teng2022sa}.
However, while bona fide data is available in abundance, spoofed data is comparatively scarce.
New techniques are needed to compensate for this data imbalance while still exploiting the volume of bona fide data needed to support reliable ASV.  We propose a multi-stage training scheme for the extraction of SASV embeddings and the use of appropriate diverse loss functions.

\begin{figure}[t]
\begin{minipage}[b]{\linewidth}
  \centering
  \centerline{\includegraphics[width=1.06\linewidth, clip]{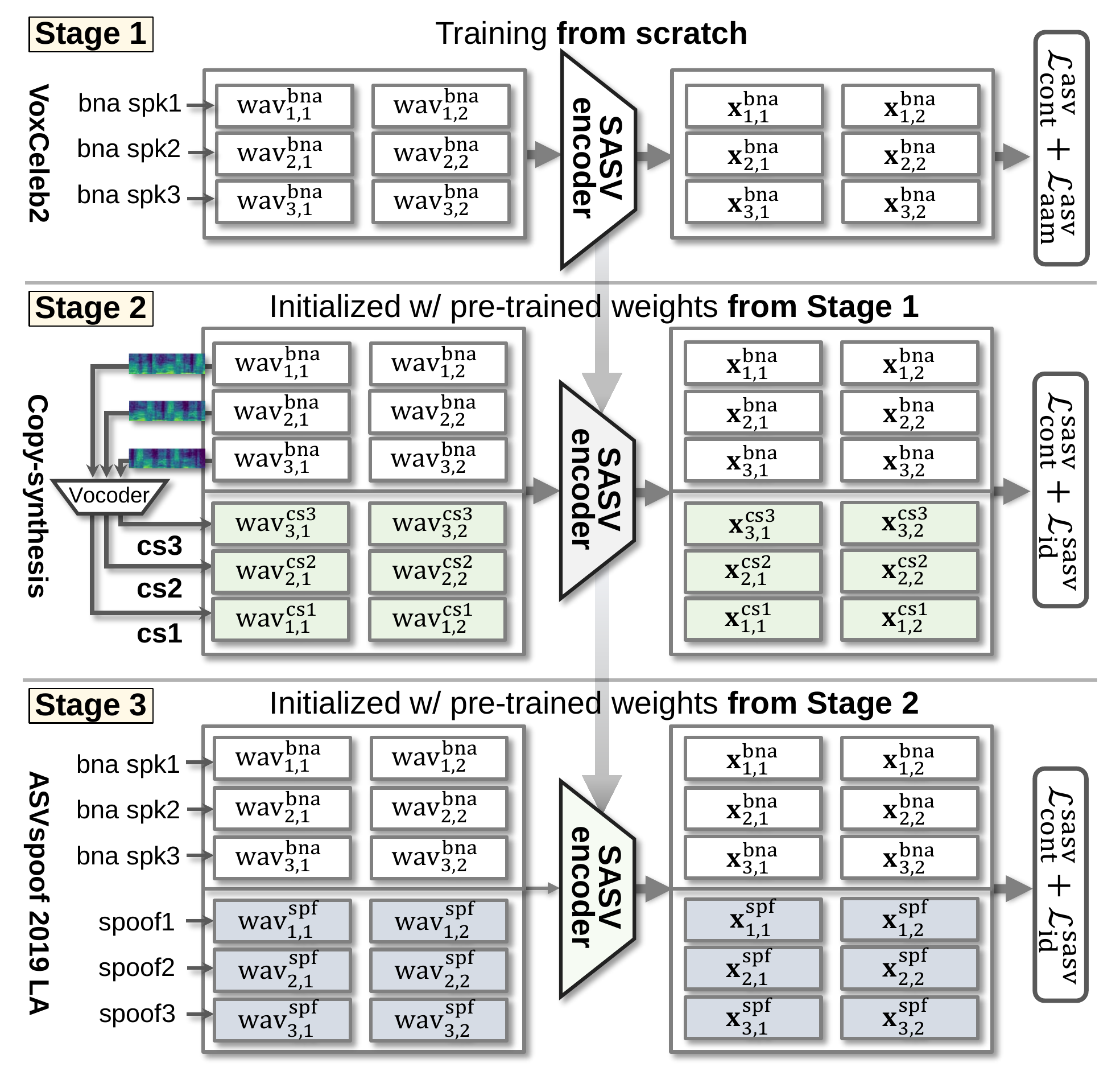}}
\vspace{-0.2cm}
\end{minipage}
\centering\caption{Proposed SASV framework with three training Stages and diverse loss functions. $\textrm{wav}^{\textrm{bna}}_{i,j}$ and $\textrm{wav}^{\textrm{spf}}_{i,j}$ denote bona fide and spoof waveform of $i$-th speaker's $j$-th utterance. CS{$k$} denotes the copy synthesis spoofed data from $k$-th speaker.}
\vspace{-14pt}
\label{fig:SASV}
\end{figure}

First, the ability to discriminate between target and bona fide non-target speakers can be learned using the VoxCeleb2 database which contains data collected from thousands of bona fide speakers.
Second, we augment the model with the ability to discriminate between bona fide and spoofed inputs by using large-scale data generated through an oracle speech synthesis system, referred to as copy synthesis~\cite{holmes1989copy}.
Third, we fine-tune the model using the ASVspoof 2019 Logical Access (LA) database which contains a mix of bona fide and spoofed utterances.
Figure~\ref{fig:SASV} illustrates the multi-training scheme.

\subsection{Stage 1: Speaker classification-based pre-training}
In Stage 1, pre-training is performed using VoxCeleb2 so that the model learns to distinguish between target and non-target trials, both of which are bona fide (i.e., conventional ASV). 
We optimize the model using the equal-weighted summation of additive angular margin (AAM) softmax loss~\cite{deng2019arcface} and contrastive loss functions. 
The combination of two functions has synergistic effects, resulting in better performance than when using either alone~\cite{heo2020clova, kwon2021ins, mun2023frequency}.

Given a mini-batch of SASV embeddings, $\textbf{x}_{i,1}$ and $\textbf{x}_{i,2}$, where $\textbf{x}_{i,k}$ denotes a SASV embedding extracted from the $k^{\text{th}}\in [1,2]$ utterance of the $i^{\text{th}} \in [1,\dots,C_{\text{spk}}]$ speaker, the contrastive loss function is defined as follows:
\begin{equation}
\mathcal{L}^{\text{asv}}_{\text{cont}} = -{\frac{1}{N}} \sum_{i}^{N} \log{\frac{\exp(f(\textbf{x}_{i,1},\textbf{x}_{i,2}))}{{\sum_{j}^{N}} \exp(f(\textbf{x}_{i,1},\textbf{x}_{j,2}))}},
\end{equation}
where $N$ is the number of speakers within a mini-batch and $f(\cdot)=\alpha \cos(\cdot)+\beta$ is a cosine similarity function between two vectors with trainable scale $\alpha$ and bias $\beta$ parameters.
Given a target speaker label $\textbf{y}_{i,k}=i\in[1,\dots,C_{\text{spk}}]$ corresponding to $\textbf{x}_{i,k}$, the AAM softmax loss is defined as:
\begin{gather}
  \mathcal{L}^{\text{asv}}_{\text{aam}} = \text{AAMsoftmax}([\textbf{x}_{i,k}, {\textbf{y}}_{i,k}]_{\forall (i,k)}; C_{\text{spk}}) \\
  = -{\frac{1}{2N}} \sum_{i,k}^{2N} \log { \frac{e^{s\cos(\theta_{\textbf{y}_{i,k}, i}+m)}}{e^{s\cos(\theta_{\textbf{y}_{i,k},i}+m)} + \sum_{j \neq \textbf{y}^{c}_{i,k}} e^{s\cos(\theta_{j, i})}}},
\end{gather}
where $s$ is a scale factor, $m$ is a margin, $\text{cos}(\theta_{j,i})$ is the normalized dot product between the weight for the $j^{\text{th}} \in [1,..,C^{\text{spk}}]$ class and the input SASV embedding $\textbf{x}_{i,k}$.
Finally, the ASV loss function for Stage~1 is defined as $\mathcal{L}^{\text{asv}} = \mathcal{L}^{\text{asv}}_{\text{cont}} + \mathcal{L}^{\text{asv}}_{\text{aam}}$.

\subsection{Stage 2: copy synthesis training with adapted SASV loss functions}
\label{sec:cs}
The training of CMs also requires large databases of both spoofed and bona fide data. 
Compared to abundant bona fide data,
generating spoofed data remains technically demanding and time-consuming.\footnote{To give an idea of the scale, the creation of the ASVspoof 2019 dataset~\cite{wang2020asvspoof} required more than 6 months of intensive effort from several contributors.}
Nevertheless, generating spoofed data with a wide variety of different algorithms is essential for generalization, namely the reliable detection performance
in the face of unseen spoofing attacks. Greater quantities, more on par with that of bona fide data, may also help to avoid the impact of data imbalance.
Moreover, no systematic study has investigated the loss function which is most appropriate for SASV, even though the training objective is critical. 
To this end, we use copy synthesis (CS) as a data augmentation to enlarge the volume of spoofed data and compare different loss functions.

\newpara{Utilization of copy synthesis data.}
Recently, an efficient means to generate massive databases for spoofing CM training was proposed in~\cite{Wang2023a}.
It uses neural network-based vocoders and a CS technique to generate spoofs from bona fide utterances using oracle acoustic features 
(e.g.\ F0 and mel spectrogram representations), and speaker embeddings (e.g.\ x-vectors)~\cite{pal2018synthetic,frank2021wavefake}. \\
Similar ideas have been reported previously~\cite{wu2013synthetic,sizov2015joint,Saratxaga2016, pal2018synthetic, frank2021wavefake, sun2023exposing}, even though most of the previous studies use traditional non-neural vocoders.
Adopting the same idea, we use the data generated via CS from VoxCeleb2 and ASVspoof 2019 LA training datasets. 
We employ the same four neural vocoders as described in~\cite{Wang2023a}: HiFiGAN~\cite{Kong2020}; WaveGlow~\cite{prenger2018waveglow}; a neural source-filter model (NSF)~\cite{wang2018neural}; a NSF model with a HiFiGAN discriminator. 
This set of vocoders includes flow, generative adversarial network (GAN), and differentiable digital signal processing techniques~\cite{engel2020ddsp}, and can generate waveforms at real-time speed, making them ideally suited to CS using large datasets. 

\newpara{Adaptive contrastive loss for SASV.}
The contrastive loss function, which is widely used in speaker verification~\cite{chung2020defence} and also in  Stage~1, maximizes speaker discrimination in the embeddings corresponding to different speakers. 
However, in the case of a spoof-aware scenario, this may affect (degrade) 
performance since 
CM labels (i.e., spoof, target bona fide) are not taken into account. 
Therefore we explore a contrastive loss function for the training of a single, integrated SASV model.
Given a mini-batch of SASV embeddings, $\textbf{x}^{c}_{i,1}$ and  $\textbf{x}^{c}_{i,2}$, where $\textbf{x}^{c}_{i,k}$ denotes a SASV embedding extracted from the $k^{\text{th}}\in [1,2]$ utterance of the $i^{\text{th}} \in [1,..,C_{\text{spk}}]$ speaker with CM label $c \in [\text{bna}, \text{spf}]\footnote{The superscript or subscript `bna' and `spf' denote bona fide and spoof data, respectively.}$,
the SASV contrastive loss is defined as follows:
\begin{gather}
  \mathcal{L}^{\text{sasv}}_{\text{cont}} = -{\frac{1}{N_{\text{bna}}}} \sum_{i}^{N_{\text{bna}}} \log {\frac{\exp(f(\textbf{x}^{\text{bna}}_{i,1},\textbf{x}^{\text{bna}}_{i,2}))} {{\sum_{j,c}^{N}} \exp(f(\textbf{x}^{\text{bna}}_{i,1},\textbf{x}^{c}_{j,2}))}},
\end{gather}
where $N_{\text{bna}}$ is the number of bona fide speakers within a mini-batch and where $N=N_{\text{bna}}+N_{\text{spf}}$ indicates the total number of bona fide and spoof speakers within a mini-batch. 
To focus on the differences in artefacts stemming from vocoding, as opposed to other sources of variation (e.g., speakers, phonetic characteristics, etc.), we include both bona fide utterances and their copy synthesized counterparts when creating mini-batches.
Given a bona fide utterance, $\textbf{x}^{\text{bna}}_{i,k}$, the corresponding CS utterance is denoted by $\textbf{x}^{\text{cs}}_{i,k}$.
Mini-batches are configured with a size of $4N_{\text{spk}}$ as follows: $\{[(\textbf{x}^{\text{bna}}_{i,k},\textbf{x}^{\text{cs}}_{i,k})]_{i,k}|\forall i\in [1,..,N_{\text{spk}}], \forall k\in [1, 2]\}$, where the vocoder type used for copy synthesis is randomly selected from the four.

\newpara{Integrated \& Multi-task SASV identification loss.} The SASV task can be considered as a multi-task problem in which there is a need to distinguish between target and non-target utterances (the traditional task of ASV), as well as bona fide and spoofed utterances (the task of CMs). 
We investigate two alternative approaches: (1)~integrated SASV identification (i.e., $C_{\text{spk}}+1$ class classification) with an additional spoof class;
(2)~multi-task SASV identification (i.e., $C_{\text{spk}}$ speaker classification and binary spoofing detection tasks).

First, bona fide and spoof SASV embeddings are defined by $\textbf{x}^{\text{bna}}_{i,k}$ and $\textbf{x}^{\text{spf}}_{i,k}$.
We set their corresponding SASV identification targets to $\hat{\textbf{y}}^{\text{bna}}_{i,k}=i\in[1,..,C_{\text{spk}}]$ and $\hat{\textbf{y}}^{\text{spf}}_{i,k}=C_{\text{spk}}+1$, respectively.
We use the AAM softmax~\cite{deng2019arcface} loss to perform $C_{\text{spk}}+1$ class identification.
For given the input SASV embeddings $\textbf{x}^{c}_{i,k}$ and their corresponding SASV identification targets $\hat{\textbf{y}}^{\text{c}}_{i,k}$, the AAM softmax is defined as:
\begin{equation}
  \mathcal{L}^{\text{sasv}}_{\text{id}_1} = \text{AAMsoftmax}([\textbf{x}^{c}_{i,k}, {\hat{\textbf{y}}}^{c}_{i,k}]_{\forall (i,k,c)}; C_{\text{spk}}+1).
\end{equation}

Second, instead of learning the SASV embeddings from an integrated single output logit, a multi-task learning~\cite{caruana1998multitask} framework can be utilized. 
Speaker and CM labels are denoted by ${\textbf{y}}^{c}_{i,k}=i\in[1,..,C_{\text{spk}}]$ and $\tilde{\textbf{y}}^{c}_{i,k}=c\in[\text{bna, spf}]$ for SASV embedding $\textbf{x}^{c}_{i,k}$, respectively.
The multi-task SASV identification loss is then given by:
\begin{gather}
 \mathcal{L}^{\text{sv}}_{\text{id}_2} = \text{AAMsoftmax}([\textbf{x}^{c}_{i,k}, {\textbf{y}}^{c}_{i,k}]_{\forall (i,k,c)}; C_{\text{spk}}), \\
  \mathcal{L}^{\text{spf}}_{\text{id}_2} = \text{AAMsoftmax}([\textbf{x}^{c}_{i,k}, \tilde{\textbf{y}}^{c}_{i,k}]_{\forall (i,k,c)}; 2), \\
  \mathcal{L}^{\text{sasv}}_{\text{id}_2} = \mathcal{L}^{\text{sv}}_{\text{id}_2} + \mathcal{L}^{\text{spf}}_{\text{id}_2},
\end{gather}
Therefore, the final loss functions in Stage 2 can be either of $\mathcal{L} = \mathcal{L}^{\text{sasv}}_{\text{cont}} + \mathcal{L}^{\text{sasv}}_{\text{id}_1}$ or $\mathcal{L} = \mathcal{L}^{\text{sasv}}_{\text{cont}} + \mathcal{L}^{\text{sasv}}_{\text{id}_2}$.  We compare the two empirically and also explore model training with only $\mathcal{L}^{\text{sasv}}_{\text{cont}}$, $\mathcal{L}^{\text{sasv}}_{\text{id}_1}$, or $\mathcal{L}^{\text{sasv}}_{\text{id}_2}$.

\subsection{Stage 3: in-domain fine-tuning}
Even though training in Stages 1 and 2 learn to discriminate bona fide non-target and spoof non-target inputs, there is a remaining domain mismatch with the evaluation protocol.
Furthermore, artefacts from the acoustic model have yet to be learned.
Hence, as a final step, we fine-tune the model using in-domain bona fide and spoofed data contained within the ASVspoof 2019 LA training partition.  
While training data corresponds to the same domain as that used for evaluation, there are differences in the spoofing algorithms used in their generation.
In Stage~3, the same loss functions of Stage~2 are used, but with different mini-batch formulation.
Specifically, and different to Stage~2, in Stage~3 we randomly sample $2N_{\text{spk}}$ bona fide (2 utterances each from $N_{\text{spk}}$ speakers) and $2N_{\text{spf}}$ spoofed utterances from the 
ASVspoof 2019 LA~\cite{wang2020asvspoof} training partition.

\section{Experimental setup}
\label{sec:exp}

\subsection{Datasets and metrics}
\label{ssec:dbs}

During the training Stages 1 and 2, we apply CS-based data augmentations using the VoxCeleb2~\cite{chung2018voxceleb2} or/and ASVspoof 2019 LA datasets~\cite{wang2020asvspoof}. 
During the training Stage 3 and evaluation phases, we only use SASV challenge data to follow the standard protocol~\cite{jung22c_INTERSPEECH}. 
Specifically, in Stage 1 training, we employ the original VoxCeleb2 dataset.
In Stage 2, we utilize either the VoxCeleb2 dataset along with the corresponding CS data, or the ASVspoof 2019 LA train and development portions, also paired with the corresponding CS data. Finally, in Stage 3, we solely rely on the ASVspoof 2019 LA train and development partitions, comprising 50,224 utterances (5,128 genuine and 45,096 spoofed) collected from 40 speakers.

Results are reported in terms of three metrics, namely the SASV-EER, the SV-EER, and the SPF-EER~\cite{jung22c_INTERSPEECH}. The SASV-EER is computed using target, non-target and spoofed trials, where only target trials should be accepted. 
The SV-EER is estimated using target and non-target trials. 
The SPF-EER is computed using target and spoofed trials; it reflects the vulnerability of the ASV system when non-target trials are replaced with spoofed trials. 

\begin{table}[t]
  \caption{ Proposed multi-Stage training. Results are reported in terms of SASV-EER (\%).
\textit{trn} and \textit{trn\&dev} refer to using train only and both train and dev set of the ASVspoof 2019 LA in Stage~3. 
  $\text{Vox}_{\text{bna}}$, $\text{Vox}_{\text{bna+cs}}$, $\text{ASV}_{\text{o}}$, $\text{ASV}_{\text{bna+cs}}$ indicate VoxCeleb2 (bona fide only), VoxCeleb2 and its CS version, original ASVspoof 2019 LA, and ASVspoof 2019 bona fide and its CS version, respectively. Rows 2, 4, and 6 do not apply Stage~3, and hence only the result after the last applied Stage is reported.}
 
  \centering
  \label{tab:main2}
  \vspace{-9.0pt}
\footnotesize{
  \renewcommand{\tabcolsep}{1.28mm}
\begin{tabularx}{\linewidth}{cccc|cc|cc}
\toprule
\multirow{2}{*}{\rotatebox{90}{\tiny rows}} & \scriptsize Stage~1     & \scriptsize Stage~2        & \scriptsize Stage~3     & \multicolumn{2}{c|}{\scriptsize SKA-TDNN} & \multicolumn{2}{c}{\scriptsize MFA-Conformer} \\
& \scriptsize \textit{pre-train}  & \scriptsize \textit{CS}      & \scriptsize \textit{fine-tune}  & \scriptsize \textit{trn}  & \scriptsize \textit{trn\&dev}  & \scriptsize \textit{trn}    & \scriptsize\textit{trn\&dev}    \\ 
\midrule
\tiny{1} &                    &                        & $\text{ASV}_\text{o}$ & 10.04           & 5.94                 & 11.47             & 7.67                   \\ 
\midrule
                    
\tiny{2} & $\text{Vox}_\text{bna}$ &                        &                     & \multicolumn{2}{c|}{16.74}                                & \multicolumn{2}{c}{20.22}                                   \\ 
\tiny{3} & $\text{Vox}_\text{bna}$ &                        & $\text{ASV}_\text{o}$ & 2.67            & \textbf{1.25}        & 2.13              & 1.49                   \\

\midrule
\tiny{4} &                    & $\text{Vox}_\text{bna+cs}$ &                     & \multicolumn{2}{c|}{13.11}                      & \multicolumn{2}{c}{14.27}\\
\tiny{5} &                    & $\text{Vox}_\text{bna+cs}$ & $\text{ASV}_\text{o}$ & 2.49            & 1.93                 & 1.91              & 1.35                   \\
\midrule

\tiny{6} & $\text{Vox}_\text{bna}$ & $\text{Vox}_\text{bna+cs}$ &                     & \multicolumn{2}{c|}{10.24}                 & \multicolumn{2}{c}{12.33}              \\
\tiny{7} & $\text{Vox}_\text{bna}$ & $\text{Vox}_\text{bna+cs}$ & $\text{ASV}_\text{o}$ & \textbf{1.83}   & 1.56                 & \textbf{1.19}     & \textbf{1.06}          \\

\midrule 
\tiny{8} &                    & $\text{ASV}_\text{bna+cs}$ &                     & 13.10            & 10.49                & 13.68             & 12.48                  \\
\tiny{9} &                    & $\text{ASV}_\text{bna+cs}$ & $\text{ASV}_\text{o}$ & 9.57            & 6.17                 & 13.46             & 10.11                  \\
                    \midrule
\tiny{10} & $\text{Vox}_\text{bna}$ & $\text{ASV}_\text{bna+cs}$ &                     & 5.62            & 4.93                 & 9.31              & 8.32                   \\
\tiny{11} & $\text{Vox}_\text{bna}$ & $\text{ASV}_\text{bna+cs}$ & $\text{ASV}_\text{o}$ & 2.48            & 1.44                 & 2.72              & 1.76                   \\
\bottomrule
\end{tabularx}
  }
  \vspace{-15pt}
\end{table}

\subsection{Implementation details}
\label{ssec:imp}
VoxCeleb2 and ASVspoof 2019 LA utterances are cropped or padded to durations of 2 and 5 seconds respectively.
Features are 80-dimensional log mel-filterbank outputs computed with a window size of 25~ms and a frame-shift of 10~ms  with mean and variance normalization.
All four vocoders described in Section~\ref{sec:cs} were trained from scratch using the VoxCeleb2 development partition and were used to generate copy synthesis data for VoxCeleb2 and the ASVspoof 2019 LA datasets. 

For all Stages, we used contrastive and AAM-softmax losses concurrently with AdamW optimizer~\cite{loshchilov2017decoupled}, except for ablation experiments.
In Stage~1, we used a batch size of 200 (2 utterances per 100 speakers), a scheduler cycle size of 25 epochs and a maximum learning rate of 0.001 which was decreased by 0.5 between each cycle. 
For Stage~2, we used a batch size of 200 (2 utterances per 50 speakers for bona fide and corresponding CS data), a scheduler cycle size of 20 epochs and a maximum learning rate of 1e-5 which was decreased by 0.5  every three cycles.
In the case of ASVspoof 2019 LA bona fide and CS data, we used a batch size of 32 (2 utterances per 8 speakers for bona fide and CS data). 
For Stage~3, we used the same settings as for ASVspoof 2019 LA data in Stage~2, and data for 16 randomly chosen spoofing attacks instead of 16 CS utterances.

\section{Results}
\label{sec:result}
\subsection{Multi-stage training and CS results}
\label{subsec:primary_result}
Results are shown in Table~\ref{tab:main2}.
The comparison of results in rows 1 and~3 shows the benefit of Stage~1 ASV pre-training when used with Stage~3 fine-tuning.
We report four types of results, using two model architectures and two Stage~3 training data configurations. The objective is to verify whether the additional use of development partition of ASVspoof2019LA in Stage~3 can be advantageous. 
SASV-EERs improved in all four columns, where it was more advantageous when stage~3 only included the train partition ({\em trn}). 
Rows 4-7 show results for Stage~2 CS-training, with or without Stage~1 and Stage~3 training.
Results in row 4 show that Stage~2 is more beneficial than Stage~1; 
performance improvements are observed 
for both models.
This makes sense since data used for Stage~1 can be viewed as {\em a subset} of the data used for Stage~2. 
The comparison of results in rows 5 and 7  show that Stage~1 training is complementary to Stage~2.

\begin{table}[t!]
  \caption{Ablation on the composition of loss functions in Stage~3, after Stage~1 (i.e., row 3 of Table~\ref{tab:main2}). {trn\&dev} partition of ASVspoof2019 LA is used for ablation study.} 
  \centering
  \renewcommand{\arraystretch}{0.9}
  \label{tab:loss_combinations}
  \renewcommand{\tabcolsep}{2.15mm} 
  \vspace{-8pt}
  \footnotesize{
  \begin{tabularx}{\linewidth}{l|rrr|rrr}
    \toprule
    \multirow{2}{*}{Loss functions}  & \multicolumn{3}{c|}{\scriptsize SKA-TDNN} & \multicolumn{3}{c}{\scriptsize MFA-Conformer}\\
                                     & \multicolumn{1}{c}{\scriptsize SASV} & \multicolumn{1}{c}{\scriptsize SV} & \multicolumn{1}{c|}{\scriptsize SPF} & \multicolumn{1}{c}{\scriptsize SASV} & \multicolumn{1}{c}{\scriptsize SV} & \multicolumn{1}{c}{\scriptsize SPF} \\

    \midrule
    $\mathcal{L}^{\text{sasv}}_{\text{cont}}$                       & 3.62  & 1.64  & 4.63   & 1.89  & 1.52  & 2.11 \\
    \midrule
    
    $\mathcal{L}^{\text{sasv}}_{\text{id}_1}$         & 1.75  & 2.35  & 1.06   & 1.96  & 2.70  & 1.01 \\         
    
    \midrule
    
    $\mathcal{L}^{\text{sasv}}_{\text{id}_2}$        & 3.57  & 5.23  & 2.03   & 1.88  & 2.85  & 0.82 \\

    \midrule
    $\mathcal{L}^{\text{sasv}}_{\text{cont}}$ $+$$\mathcal{L}^{\text{sasv}}_{\text{id}_1}$    & \textbf{1.25} & \textbf{1.27}  & \textbf{1.23}   & \textbf{1.49} & \textbf{1.79}  & \textbf{1.28} \\       
   
    \midrule
    
    $\mathcal{L}^{\text{sasv}}_{\text{cont}}$$+$$\mathcal{L}^{\text{sasv}}_{\text{id}_2}$     & 3.00 & 3.25  & 2.81     & 2.39 & 3.00  & 1.81 \\

    \bottomrule
  \end{tabularx}
  }
  \vspace{-17pt}
\end{table}

Rows 8 to 11 show results using Stage~2 training with ASVspoof 2019 LA CS instead of VoxCeleb2 CS. 
The comparison of results in rows 8 and 9 with those in rows 4 and 5 shows similar results with no fine-tuning, but worse results with; 
fine tuning using VoxCeleb, which contains data collected from a far greater number of speakers, is the most beneficial. 
The comparison of results in rows 10 and 11 to rows 8 and 9 show slightly improved performance when ASVspoof 2019 LA data is used in Stage~2 and combined with VoxCeleb data in Stage~1.
Even so, results are slightly worse than corresponding results in rows 4-7 for which VoxCeleb data is used for Stage 2. 
These results show that the use of data collected from a larger number of speakers leads to better speaker discrimination and SASV performance.

Results for the two models show similar trends. 
The use of all three stages together with VoxCeleb data for Stage 2 (row 7), results in the lowest SASV-EER for both models; the only exception is for the SKA-TDNN model fine-tuned in Stage~3 with both train and development sets. 
The best result of 1.06\% is produced with 
the MFA-Conformer model when trained using all three stages and when fine-tuned using both train and development sets of the ASVspoof 2019 LA database.
In summary, Stage~1 pre-training with VoxCeleb data, followed by Stage~2 training with CS data, and then fine-tuning with in-domain ASVspoof 2019 LA data generally leads to the best performance.

\begin{table}[t!]
  \caption{Comparison with previous works in the literature. With Stage 1 and Stages 1-2 denote the results with pre-training (Stage 1) and sequential CS-training (Stages 1-2), respectively.}
  \centering
  \renewcommand{\arraystretch}{0.99}
  \label{tab:comparison_others}
  \renewcommand{\tabcolsep}{1.50mm} 
  \vspace{-8pt}
  \footnotesize{
  \begin{tabularx}{\linewidth}{lcrrr}
    \toprule
    System & Type & SASV & SV & SPF\\
    \midrule
    ECAPA-TDNN~\cite{desplanques2020ecapa} & ASV & 23.83 & 1.63  & 30.75 \\
    AASIST~\cite{jung2022aasist} & CM       & 24.38 & 49.24 & 0.67 \\

    \midrule
SASV Baseline~\cite{shim2022baseline}           & Score-level Fusion       & 1.71&1.66  & 1.76 \\
  \cite{hyu_sasv_2022} & Emb. level fusion& 0.28&0.28&0.28\\ 
    \cite{ge2022joint}& Joint-optimization& 1.53&2.44&0.75\\
   \cite{teng2022sa}     & Single system    & 4.86  & 8.06  & 0.50 \\

    \midrule
    \textbf{Ours - \textit{Stage 3 only}}        & Single system    & 11.47 & 13.46 & 10.30 \\
    \>\> {\it with Stage~1}         & Single system    & 2.13  & 3.43  & 1.08  \\
    \>\> {\it with Stages~1-2}     & Single system    & \textbf{1.19} & \textbf{1.83} & \textbf{0.58} \\
    
    \bottomrule
  \end{tabularx}
  \vspace{-17pt}
  }
\end{table}
\subsection{Loss function comparison}
\label{subsec:loss_result}
Table~\ref{tab:loss_combinations} shows 
the results adopting Stages~1 and~3, with different loss functions and their combination.
When using only one of the three loss functions, the trends differ for each  model.
The use of contrastive loss ($\mathcal{L}^{\text{sasv}}_{\text{cont}}$) with integrated identification loss ($\mathcal{L}^{\text{sasv}}_{\text{id}_1}$), leads to the best performance for both models. 
The comparison of identification losses ($\mathcal{L}^{\text{sasv}}_{\text{id}_1}$ and $\mathcal{L}^{\text{sasv}}_{\text{id}_2}$) shows that
integrated outperformed its counterpart in most cases.

\subsection{Performance comparison}
\label{subsec:comparison_result}
Table~\ref{tab:comparison_others} shows a comparison of results for the proposed single SASV system to those for competing  systems reported in the literature. 
The first two rows show results for the individual ASV and CM subsystems used by SASV challenge baselines~\cite{jung22c_INTERSPEECH}. 
Rows 3 to 6 show results for the SASV B1 baseline\footnote{The 2022 SASV challenge B1 baseline with softmax applied to the CM output which improves results notably.}, 
the best embedding-level fusion system,
the best jointly-optimized system, and best single end-to-end system reported in the literature.\footnote{Better results are reported in the literature, but these correspond to \emph{ensemble} systems.  The comparisons presented here are restricted to systems which employ at most one CM and one ASV subsystem.} 
Results for our best system (bottom row) correspond to an 84\% relative improvement in SASV-EER compared to the best single system for which the SASV-EER is 4.86\%. 
This pushes the state-of-the-art in single SASV approaches dramatically.
It also outperforms the best jointly-optimized solution. 

\section{Conclusions}

We present a new integrated SASV embedding extractor that achieves an SASV-EER of 1.06\%, significantly surpassing previous best results for such systems. This approach bridges the performance gap between single integrated SASV systems and fusion-based methods that combine pre-trained ASV and CM models. Our multi-stage training solution benefits from pre-training with large amount of data also with CS, which addresses the scarcity of spoofed data and the imbalance with genuine training data. 

\vspace{-5pt}
\section{Acknowledgements}
This work is partially supported by the COMPA grant funded by the Korea government (MSIT and Police) (No.~RS-2023-00235082), Academy of Finland (Decision No.~349605, project ``SPEECHFAKES''), JST CREST (JPMJCR18A6, JPMJCR20D3), and MEXT KAKENHI (21K17775, 21H04906).

\clearpage
\bibliographystyle{IEEEtran}
\bibliography{shortstrings,mybib}

\end{document}